\documentclass[
    ,final            
  ]
  {aipproc}
\usepackage{amssymb}
\layoutstyle{8x11single}
\begin{document}
\title{Weak-universal critical behavior and quantum critical point of the exactly soluble 
spin-1/2 Ising-Heisenberg model with the pair $XYZ$ Heisenberg and quartic Ising interactions}


\author{Jozef Stre\v{c}ka}{address={Department of Theoretical Physics and Astrophysics, 
Faculty of Science, \\ P. J. \v{S}af\'{a}rik University, Park Angelinum 9, 040 01 Ko\v{s}ice, 
Slovak Republic}}
\author{Lucia \v{C}anov\'a}{address={Department of Applied Mathematics, Faculty of Mechanical Engineering, \\ Technical University, Letn\'a 9, 042 00 Ko\v{s}ice, Slovak Republic}}
\author{Kazuhiko Minami}{address={Graduate School of Mathematics, Nagoya University,                  
             Nagoya 464-8602, Japan}}    

\begin{abstract}
Spin-1/2 Ising-Heisenberg model with $XYZ$ Heisenberg pair interaction and two different 
Ising quartic interactions is exactly solved with the help of the generalized star-square transformation, which establishes a precise mapping equivalence with the corresponding 
eight-vertex model on a square lattice generally satisfying Baxter's zero-field (symmetric) condition. 
The investigated model exhibits a remarkable weak-universal critical behavior with two marked 
wings of critical lines along which critical exponents vary continuously with the interaction 
parameters. Both wings of critical lines merge together at a very special quantum critical point 
of the infinite order, which can be characterized through diverging critical exponents. 
The possibility of observing reentrant phase transitions in a close vicinity of the quantum 
critical point is related to a relative strength of the exchange anisotropy in the $XYZ$ Heisenberg 
pair interaction. 
\end{abstract}

\maketitle

\section{Introduction}

Exactly soluble quantum spin models belong to the most fascinating topics to deal with in 
the area of modern equilibrium statistical mechanics \cite{baxt82,matt93,wu08}. 
It should be pointed out, however, that quantum effects usually compete with a cooperative nature 
of spontaneous long-range ordering and thus, it is quite intricate to find an exactly solvable 
model that simultaneously exhibits both spontaneous long-range order as well as obvious 
macroscopic features of quantum origin. On the other hand, it is a competition between quantum 
and cooperative phenomena that is an essential ingredient for observing a quite remarkable 
and unexpected behavior of low-dimensional quantum spin models.   

The hybrid Ising-Heisenberg models on decorated planar lattices, whose nodal sites are occupied 
by the classical Ising spins and decorating sites by the quantum Heisenberg ones, belong to 
the simplest rigorously solved quantum spin models that exhibit a spontaneous long-range ordering 
with apparent quantum manifestations. It is worthwhile to remark, moreover, that the 
Ising-Heisenberg planar models \cite{stre02,jasc02,stre04,jasc04,stre06,cano07,stre08,yao08,cano08,jasc08} where a finite cluster 
of the Heisenberg spins interacts with either two or three nodal Ising spins are in principle 
tractable by the use of generalized decoration-iteration or star-triangle transformations \cite{fish59,syoz72,roja08}, which establish a precise mapping equivalence between them 
and the spin-1/2 Ising model on the corresponding undecorated planar lattice \cite{baxt82,syoz72,lavi99,utiy51,domb60,lin86}. Among other matters, the exact solutions 
for this special class of the Ising-Heisenberg planar models might serve in evidence that 
these rigorously solvable models exhibit a strong-universal critical behavior, which can be  characterized by critical exponents from the standard Ising universality class. Contrary to this, 
the more interesting weak-universal behavior \cite{suzu74} of the critical exponents has 
been recently announced for two Ising-Heisenberg planar models \cite{valv09,stre09}, 
where a finite cluster of the Heisenberg spins interacts with four nodal Ising spins. 
The spin-1/2 Ising-Heisenberg model with the pair $XYZ$ Heisenberg interaction and 
two quartic Ising interactions \cite{stre09} has surprisingly turned out to be the fully 
exactly solvable model due to a validity of the precise mapping equivalence with Baxter's 
zero-field (symmetric) eight-vertex model \cite{baxt82,baxt71,baxt72}. The main purpose of 
this work is to examine in detail how the weak-universal critical behavior of this exactly 
soluble model depends on a spatial anisotropy in two quartic Ising interactions and 
on the exchange anisotropy of the $XYZ$ Heisenberg pair interaction, whose effect have 
not been dealt with in our preceding work \cite{stre09} for the most general case. 

This paper is so organized. In the following section, we will describe the hybrid Ising-Heisenberg model and recall basic steps of the exact mapping procedure to the zero-field eight-vertex model. The most interesting results for the ground-state and finite-temperature phase diagrams, which are supported by a detailed analysis of critical exponents, are subsequently presented 
in the next section. Finally, some concluding remarks are mentioned along 
with a brief summary of the most important scientific achievements in the last section.

\section{Ising-Heisenberg model and its exact mapping equivalence with 
the zero-field eight-vertex model}

\begin{figure}
\vspace{0cm}
\includegraphics[width=0.35\textwidth]{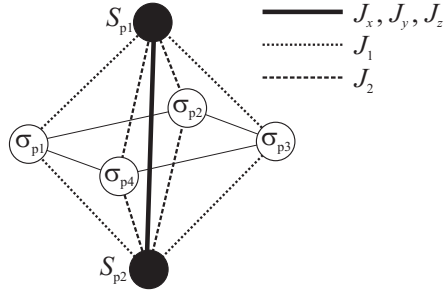}
\vspace{0.0cm}
\caption{The elementary unit cell of the spin-1/2 Ising-Heisenberg model. Full (empty) circles
denote positions of the Heisenberg (Ising) spins, thick solid line represents the pairwise $XYZ$ Heisenberg interaction between the apical Heisenberg spins and both types of broken lines connect 
spins involved in the quartic Ising interactions. Thin solid lines connecting four Ising spins 
are guide for eyes only.}
\label{fig1}
\end{figure}

Let us consider a two-dimensional lattice of edge-sharing octahedrons. Figure~\ref{fig1} 
schematically displays the elementary unit cell of the two-dimensional  lattice, 
i.e. an octahedron, which contains four Ising spins $\sigma=1/2$ in its basal plane 
and two Heisenberg spins $S=1/2$ in its apical positions. Let each edge of the octahedron, which connects two Ising spins, be a common edge of two adjacent octahedrons so that an ensemble of 
all Ising spins will form a square lattice and the Heisenberg spins will be located above and 
below a center of each elementary square face. Suppose furthermore that both apical Heisenberg spins interact together via the pair $XYZ$ Heisenberg interaction,
while they also take part in two different quartic Ising-type interactions with two Ising spins 
from opposite corners of a square face (see figure~\ref{fig1}). For further convenience, 
the total Hamiltonian can be written as a sum over all elementary unit cells (octahedrons) 
$\hat{{\cal H}} = \sum_{p} \hat{{\cal H}}_p$, where each cluster Hamiltonian 
$\hat{{\cal H}}_p$ contains one pair interaction between the apical Heisenberg spins and 
two quartic Ising-type interactions between the Heisenberg spins and their four Ising neighbors 
\begin{eqnarray}
\hat{{\cal H}}_p = - \left(J_x \hat{S}_{p1}^x \hat{S}_{p2}^x + J_y \hat{S}_{p1}^y \hat{S}_{p2}^y 
                                  + J_z \hat{S}_{p1}^z \hat{S}_{p2}^z \right) 
  - J_1 \hat{S}_{p1}^z \hat{S}_{p2}^z \hat{\sigma}_{p1}^z \hat{\sigma}_{p3}^z 
            - J_2 \hat{S}_{p1}^z \hat{S}_{p2}^z \hat{\sigma}_{p2}^z \hat{\sigma}_{p4}^z. 
\label{ham}	   
\end{eqnarray}
Above, the interaction parameters $J_x, J_y, J_z$ label spatial components of the anisotropic 
$XYZ$ interaction between the Heisenberg spins, while the interaction parameters $J_1$ and $J_2$ 
label two quartic Ising-type interactions between both apical Heisenberg spins and two Ising spins 
from opposite corners of a square face along two different diagonal directions 
(see figure~\ref{fig1}).

The crucial step of our calculation lies in an evaluation of the partition function. A validity 
of the commutation relation $[\hat{{\cal H}}_i, \hat{{\cal H}}_j] = 0$ between different octahedron-cluster Hamiltonians allows a partial factorization of the partition function into 
the following product 
\begin{eqnarray}
{\mathcal Z}_{\rm IHM} = \sum_{\{\sigma \}} \prod_{p} \mbox{Tr}_p \exp(- \beta \hat{{\cal H}}_p)
 = \sum_{\{\sigma \}} \prod_{p} \omega_p (\sigma_{p1}^z, \sigma_{p2}^z, \sigma_{p3}^z, \sigma_{p4}^z),  
\label{pf1}
\end{eqnarray}
where $\beta = 1/(k_{\rm B} T)$, $k_{\rm B}$ is Boltzmann's constant and $T$ is the absolute 
temperature. The summation $\sum_{\{\sigma \}}$ to emerge in Eq.~(\ref{pf1}) is carried out 
over all possible configurations of the Ising spins, the product runs over all octahedron unit 
cells and the symbol $\mbox{Tr}_p$ stands for a trace over spin degrees of freedom of the 
Heisenberg spin pair from the $p$th octahedron. In the latter step of our calculation we have 
used a straightforward diagonalization of the Hamiltonian (\ref{ham}) of the $p$th octahedron 
in order to obtain the relevant trace over spin degrees of freedom of the Heisenberg spin pair.
This procedure yields some effective Boltzmann's weight $\omega_p$, which explicitly depends 
merely on four Ising spins $\sigma_{p1}$, $\sigma_{p2}$, $\sigma_{p3}$ and $\sigma_{p4}$ from 
the basal plane of the $p$th octahedron. In addition, the explicit form of the effective Boltzmann's 
factor $\omega_p$ immediately implies a possibility of performing the generalized star-square
transformation
\begin{eqnarray}
\omega_p (\sigma_{p1}^z, \sigma_{p2}^z, \sigma_{p3}^z, \sigma_{p4}^z) \! \! \! &=& \! \! \!   
 2 \exp \left[\frac{\beta}{4} \left( J_z + J_1 \sigma_{p1}^z \sigma_{p3}^z + J_2 \sigma_{p2}^z \sigma_{p4}^z \right) \right] \cosh \left[ \frac{\beta}{4} \left(J_x - J_y \right) \right] \nonumber \\
  \! \! \! &+&  \! \! \!
       2 \exp \left[- \frac{\beta}{4} \left( J_z + J_1 \sigma_{p1}^z \sigma_{p3}^z + J_2 \sigma_{p2}^z \sigma_{p4}^z \right) \right] \cosh \left[ \frac{\beta}{4} \left(J_x + J_y \right) \right]
       \nonumber \\ \! \! \! &=& \! \! \!
R_0 \exp(\beta R_1 \sigma^z_{p1} \sigma^z_{p3} + \beta R_2 \sigma^z_{p2} \sigma^z_{p4} 
       + \beta R_4 \sigma^z_{p1} \sigma^z_{p2} \sigma^z_{p3} \sigma^z_{p4}),
\label{sst} 
\end{eqnarray}
which substitutes the effective Boltzmann's weight $\omega_p$ by the equivalent expression containing two pair ($R_1$ and $R_2$) and one quartic ($R_4$) interaction between four nodal Ising spins from an elementary square face of the square lattice. Of course, the algebraic transformation (\ref{sst}) must satisfy the 'self-consistency' condition, which means that it must hold independently of spin states 
of four Ising spins involved therein. It can be easily verified that a substitution of all sixteen possible spin configurations of the nodal Ising spins gives just four independent equations, 
which unambiguously determine so far not specified mapping parameters $R_0$, $R_1$, $R_2$, and $R_4$ 
\begin{eqnarray}
R_0 =  \left(\omega_1 \omega_3 \omega_5 \omega_7 \right)^{1/4},   \quad 
\beta R_1 =  \ln \left( \frac{\omega_1 \omega_7}{\omega_3 \omega_5} \right), \quad 
\beta R_2 =  \ln \left( \frac{\omega_1 \omega_5}{\omega_3 \omega_7} \right),  \quad
\beta R_4 = 4 \ln \left( \frac{\omega_1 \omega_3}{\omega_5 \omega_7} \right),
\label{ev}
\end{eqnarray}  
that can be expressed in terms of four different Boltzmann's weights $\omega_i$ ($i = 1,3,5,7$) 
\begin{eqnarray}
\omega_1 (+, \pm, +, \pm) \! \! \! \! \! \! \! &=& \! \! \! \! \! \! \!
2 \exp \! \left[ \frac{\beta}{4} \left( J_z + \frac{J_1 + J_2}{4} \right) \right] \!
  \cosh \! \left[ \frac{\beta}{4} \left(J_x - J_y \right) \right]
      + 2 \exp \! \left[- \frac{\beta}{4} \left( J_z + \frac{J_1 + J_2}{4} \right) \right] \!
  \cosh \! \left[ \frac{\beta}{4} \left(J_x + J_y \right) \right]\! \!, \label{bw1} \\
\omega_3 (+, \mp, -, \pm) \! \! \! \! \! \! \! &=& \! \! \! \! \! \! \!
2 \exp \! \left[ \frac{\beta}{4} \left( J_z - \frac{J_1 + J_2}{4} \right) \right] \!
  \cosh \! \left[ \frac{\beta}{4} \left(J_x - J_y \right) \right]
     +  2 \exp \! \left[- \frac{\beta}{4} \left( J_z - \frac{J_1 + J_2}{4} \right) \right] \!
  \cosh \! \left[ \frac{\beta}{4} \left(J_x + J_y \right) \right]\! \!,  \label{bw3} \\
\omega_5 (\pm, +, \mp, +) \! \! \! \! \! \! \! &=& \! \! \! \! \! \! \!
2 \exp \! \left[ \frac{\beta}{4} \left( J_z - \frac{J_1 - J_2}{4} \right) \right] \!
  \cosh \! \left[ \frac{\beta}{4} \left(J_x - J_y \right) \right]
     +  2 \exp \! \left[- \frac{\beta}{4} \left( J_z - \frac{J_1 - J_2}{4} \right) \right] \!
  \cosh \! \left[ \frac{\beta}{4} \left(J_x + J_y \right) \right]\! \!,  \label{bw5} \\
\omega_7 (+, \mp, +, \pm) \! \! \! \! \! \! \! &=& \! \! \! \! \! \! \!
2 \exp \! \left[ \frac{\beta}{4} \left( J_z + \frac{J_1 - J_2}{4} \right) \right] \!
  \cosh \! \left[ \frac{\beta}{4} \left(J_x - J_y \right) \right]
     +  2 \exp \!\left[- \frac{\beta}{4} \left( J_z + \frac{J_1 - J_2}{4} \right) \right] \!
  \cosh \! \left[ \frac{\beta}{4} \left(J_x + J_y \right) \right]\! \!. \label{bw7}
\end{eqnarray}   
Note that the effective Boltzmann's weights (\ref{bw1})-(\ref{bw7}) are assigned to eight spin configurations of four nodal Ising spins explicitly specified in round brackets (the symbol 
$\pm$ denotes the spin state $\sigma_{p \alpha}^z = \pm 1/2$, $\alpha=1,2,3,4$), as well as,
another eight spin configurations, which can be obtained from them under the reversal of all 
four nodal Ising spins.

A substitution of the generalized star-square transformation (\ref{sst}) with appropriately chosen 
mapping parameters (\ref{ev}) into Eq.~(\ref{pf1}) then straightforwardly leads to a precise mapping equivalence between the spin-1/2 Ising-Heisenberg model and the zero-field eight-vertex model. 
As a result of this procedure, one actually obtains a simple mapping relationship 
\begin{eqnarray}
{\mathcal Z}_{\rm IHM} (\beta, J_x, J_y, J_z, J_1, J_2) 
= R_0^{2N} {\mathcal Z}_{8-v} (\beta, R_1, R_2, R_4),   
\label{pf2}
\end{eqnarray}
which connects the partition function of the investigated Ising-Heisenberg model with 
the partition function of the spin-1/2 Ising model on two interpenetrating square lattices 
that are coupled together by means of the quartic interaction ($N$ is the total number of 
the Ising spins). It should be mentioned that the latter model is nothing but one 
of many alternative definitions of \emph{the zero-field eight-vertex model}  \cite{baxt82,baxt71,baxt72}, which is reformulated in the Ising spin representation
following the ideas of Wu \cite{wu71}, Kadanoff and Wegner \cite{kada71}. Besides, 
the physical meaning of the mapping parameters $R_1$, $R_2$ and $R_4$ becomes quite evident
from the mapping transformation (\ref{sst}). The parameters $R_1$ and $R_2$ denote the 
effective pair interactions in two different interpenetrating Ising square lattices and 
the mapping parameter $R_4$ determines the effective quartic interaction that couples together
both Ising square lattices. Last but not least, the mapping parameter $R_0$ (multiplicative 
factor in Eq.~(\ref{pf2})) in fact represents the partition function of the Heisenberg 
spin pair in the effective field produced by the four enclosing nodal Ising spins.

It becomes quite clear from the mapping relationship (\ref{pf2}) that the Ising-Heisenberg model becomes critical just if the corresponding zero-field eight-vertex model becomes critical as well. 
As a matter of fact, the partition function ${\mathcal Z}_{\rm IHM}$ will exhibit a non-analyticity 
if and only if the corresponding partition function ${\mathcal Z}_{8-v}$ will exhibit the same 
kind of the non-analytic behavior, because the parameter $R_0$ is analytic in the whole region
of interaction parameters. In this regard, the respective critical points of the Ising-Heisenberg 
model can be found from the relevant critical condition of the zero-field eight-vertex model \cite{baxt82,baxt71,baxt72}
\begin{eqnarray}
\omega_1 + \omega_3 + \omega_5 + \omega_7 = 2 \mbox{max} \{\omega_1, \omega_3, \omega_5, \omega_7 \},
\label{cc}
\end{eqnarray}   
which determines phase transitions of the Ising-Heisenberg model on assumption that the effective Boltzmann's weights (\ref{bw1})-(\ref{bw7}) are substituted into this critical condition. 
It should be also mentioned that the critical exponents of the zero-field eight-vertex model \cite{baxt82,baxt71,baxt72} continuously change with the parameter $\mu = 2 \arctan(\omega_5 \omega_7/ \omega_1 \omega_3)^{1/2}$ by following the formulas
\begin{eqnarray}
\alpha = \alpha' = 2 - \frac{\pi}{\mu}, \qquad \beta = \frac{\pi}{16 \mu}, \qquad 
\nu = \nu' = \frac{\pi}{2 \mu}, \qquad \gamma = \gamma' = \frac{7 \pi}{8 \mu}, \qquad \delta = 15, 
\qquad \eta = \frac{1}{4}.
\label{ce}
\end{eqnarray} 
Consequently, the relations (\ref{ce}) will also govern changes of the critical exponents 
of the Ising-Heisenberg model when the effective Boltzmann's weights (\ref{bw1})-(\ref{bw7}) 
are used for a calculation of the parameter $\mu$. 

\section{Results and discussion}

In this part, let us proceed to a discussion of the most interesting results obtained 
for critical properties of the spin-1/2 Ising-Heisenberg model with the pair and quartic interactions.
Before doing so, however, it is worthy of notice that some special cases of this model system
have already been investigated by the present authors in our earlier paper \cite{stre09}, 
where the critical behavior of the particular case with two identical quartic interactions 
$J_1 = J_2$ and the more symmetric $XXZ$ Heisenberg interaction ($J_x = J_y \neq J_z$) 
was explored in detail by assuming both ferromagnetic as well as antiferromagnetic pair interaction. 
Exact results for this more symmetric version of the Ising-Heisenberg model indicate that 
the model with the antiferromagnetic pair interaction exhibits less significant changes 
of both critical temperatures as well as critical exponents than the model with the 
ferromagnetic pair interaction. As a matter of fact, it has been demonstrated that 
only the Ising-Heisenberg model with the ferromagnetic pair interaction shows a unusual 
quantum critical point of the infinite order, which characterizes a remarkable singular 
behavior of the critical exponents near the isotropic limit of the Heisenberg pair interaction 
($J_x = J_y = J_z$). 
With this background, the main purpose of the present work is to shed light on how 
this strange weak-universal critical behavior will change by introducing a spatial anisotropy 
in two quartic Ising-type interactions ($J_1 \neq J_2$) or by assuming the most 
general form of the $XYZ$ exchange anisotropy in the Heisenberg pair interaction. 

It is worthwhile to remark that the investigated Ising-Heisenberg model possesses 
a rather high symmetry, because all four different Boltzmann's weights (\ref{bw1})-(\ref{bw7}) 
are mutually interchangeable under the transformations $J_1 \to -J_1$ and/or $J_2 \to -J_2$. 
This means, among other matters, 
that one may further restrict both quartic interaction parameters to positive values, since 
the transformations $J_1 \to -J_1$ and $J_2 \to -J_2$ merely cause rather trivial changes of 
the nodal Ising spins $(\sigma_{p1}^z, \sigma_{p2}^z, \sigma_{p3}^z, \sigma_{p4}^z) \to (\pm \sigma_{p1}^z, \sigma_{p2}^z, \mp \sigma_{p3}^z, \sigma_{p4}^z)$ and $(\sigma_{p1}^z, \sigma_{p2}^z, \sigma_{p3}^z, \sigma_{p4}^z) \to (\sigma_{p1}^z, \pm \sigma_{p2}^z, \sigma_{p3}^z, \mp \sigma_{p4}^z)$, respectively. Owing to this fact, let us further assume that all the interaction 
terms $J_x$, $J_y$, $J_z$, $J_1$, and $J_2$ entering the Hamiltonian (\ref{ham}) are positive 
and moreover, the symmetry of the Hamiltonian allows us to consider $J_x \geq J_y$ and $J_1 \geq J_2$ 
without loss of the generality. For easy reference, the number of free parameters is 
lowered by introducing the following set of dimensionless parameters: $k_{\rm B}T/J_z$ 
marks the dimensionless temperature, $J_1/J_z$ and $J_2/J_z$ are proportional to a relative 
strength of the quartic Ising-type interactions, and finally, the parameters $J_x/J_z$ 
and $J_y/J_x$ measure a relative strength of the exchange anisotropy in the $XYZ$ Heisenberg 
pair interaction. The former anisotropy parameter $J_x/J_z$ determines a difference in 
the exchange interactions along a quantization $z$-axis and its perpendicular $x$-axis, 
whereas the latter anisotropy parameter determines a difference between the stronger 
and weaker exchange interaction in the $xy$-plane.   

First, let us take a closer look at the ground-state behavior. It can be readily understood 
that the ground-state spin arrangement will always correspond to the lowest-energy eigenstate 
that enters into the greatest Boltzmann's weight among the four Boltzmann's weights given by Eqs.~(\ref{bw1})-(\ref{bw7}). In the zero temperature limit, the greatest Boltzmann's weight
is $\omega_1$ if $J_y < J_z$ or $\omega_3$ if $J_y > J_z$. This observation would suggest 
that the ground-state spin arrangement will basically change whenever the weaker exchange 
interaction $J_y$ in the $xy$-plane exceeds the exchange interaction $J_z$ along the quantization axis. 
One actually finds that the lowest-energy eigenstate is either
\begin{eqnarray}
|{\rm I} \rangle \! \! \! &=& \! \! \! \prod_p |+,\pm,+,\pm \rangle_{\sigma_p} 
\frac{1}{\sqrt{2}} \left(|+,+ \rangle + |-,- \rangle \right)_{S_p}, 
\label{gsa} 
\end{eqnarray} 
if $J_y < J_z$, or, 
\begin{eqnarray}
|{\rm II} \rangle \! \! \! &=& \! \! \! \prod_p |+,\pm,-,\mp \rangle_{\sigma_p} 
\frac{1}{\sqrt{2}} \left(|+,- \rangle + |-,+ \rangle \right)_{S_p}, 
\label{gsb} 
\end{eqnarray} 
if the reverse inequality holds. In Eqs.~(\ref{gsa})--(\ref{gsb}), the former ket vector unambiguously determines the states of four nodal Ising spins from an elementary square face of the $p$th octahedron
and the latter ket vector unambiguously specifies the relevant state of the Heisenberg spin pair. 
It should be also noticed that another equivalent representations of these eigenstates can be obtained from the eigenvectors (\ref{gsa})--(\ref{gsb}) under the reversal of all four nodal Ising spins and consequently, the phases $|{\rm I} \rangle$ and $|{\rm II} \rangle$ are both four-fold degenerate. 

Let us now make a few comments on spin arrangements appearing in both ground-state phases. 
In the phase $|{\rm I} \rangle$, the Ising spins placed on a square lattice either exhibit 
a perfect ferromagnetic or antiferromagnetic long-range order, which is accompanied with 
the entangled spin state $\left(|+,+ \rangle + |-,- \rangle \right)/\sqrt{2}$ consisting 
of both ferromagnetic states of the Heisenberg spin pairs. On the other hand, the Heisenberg 
spin pairs reside in the phase $|{\rm II} \rangle$ the entangled spin state 
$\left(|+,- \rangle + |-,+ \rangle \right)/\sqrt{2}$ consisting of both antiferromagnetic 
states and the Ising spins prefer a superantiferromagnetic long-range order with 
the ferromagnetic alignment in a horizontal direction and the antiferromagnetic 
alignment in a vertical direction, or vice versa. To compare with, is might quite useful 
to mention that the phase $|{\rm II} \rangle$ preserves its character even if the less general 
case of the $XXZ$ exchange anisotropy is assumed, whereas the spin arrangement of the phase 
$|{\rm I} \rangle$ drastically changes disentanglement of both ferromagnetic states 
of the Heisenberg spin pairs
\begin{eqnarray}
|{\rm I}' \rangle = \prod_p |+,\pm,+,\pm \rangle_{\sigma_p} |\pm, \pm \rangle_{S_p}. 
\label{gsaa} 
\end{eqnarray} 
In the particular limit $J_x = J_y$, the phase $|{\rm I}' \rangle$ actually becomes macroscopically 
degenerate as the Heisenberg spin pairs may choose independently of each other one of two 
ferromagnetic states $|\pm, \pm \rangle_{S_p}$. Hence, it follows that the phase $|{\rm I}' \rangle$ 
will exhibit a rather high macroscopical degeneracy of the order $4 \times 2^{N}$, which is proportional to the total number of the Heisenberg spin pairs (remember that $N$ simultaneously 
marks the total number of the Ising spins, the total number of the Heisenberg spin pairs, 
as well as, the total number of elementary unit cells).
 
Now, we will turn our attention to a detailed study of finite-temperature phase diagrams. 
One should recall that phase transition lines of the spin-1/2 Ising-Heisenberg model with 
the pair and quartic interactions can be straightforwardly calculated from the critical condition 
of the corresponding zero-field eight-vertex model by substituting the effective Boltzmann's weights (\ref{bw1})--(\ref{bw7}) into Eq.~(\ref{cc}). It is worthwhile to remember, moreover, 
that the greatest Boltzmann's weight is either $\omega_1$ or $\omega_3$. For both particular 
cases, the critical condition can uniquely be expressed as 
\begin{eqnarray}
\frac{\sinh \left[ \frac{\beta_{\rm c}}{16} \left(J_1 + J_2 \right) \right]}
{\cosh \left[ \frac{\beta_{\rm c} }{16} \left(J_1 - J_2 \right) \right]} 
= \pm \frac{\cosh \left[ \frac{\beta_{\rm c} }{4} \left(J_x - J_y \right) \right] 
+ \exp \left(- \frac{\beta_{\rm c} J_z}{2} \right) \cosh \left[ \frac{\beta_{\rm c}}{4} 
\left(J_x + J_y \right) \right]}{\cosh \left[ \frac{\beta_{\rm c} }{4} \left(J_x - J_y \right) \right] 
- \exp \left(- \frac{\beta_{\rm c} J_z}{2} \right) \cosh \left[ \frac{\beta_{\rm c}}{4} 
\left(J_x + J_y \right) \right]},
\label{ccef}
\end{eqnarray} 
where $\beta_{\rm c} = 1/(k_{\rm B} T_{\rm c})$, $T_{\rm c}$ is the critical temperature
and the plus (minus) sign applies for a situation when $\omega_1 > \omega_3$ ($\omega_1 < \omega_3$). 

\begin{figure}
\includegraphics[width=0.5\textwidth]{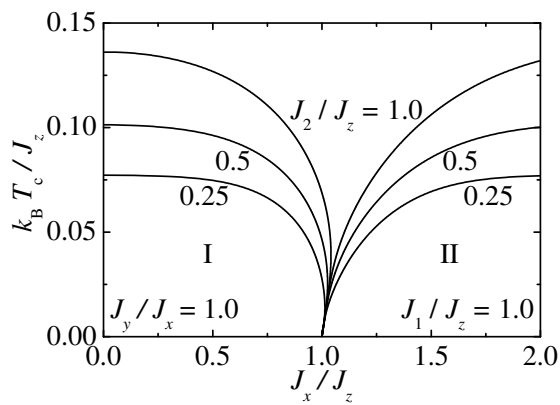}
\vspace{0.0cm}
\caption{The dimensionless critical temperature as a function of the anisotropy parameter $J_x/J_z$
for the Ising-Heisenberg model with the $XXZ$ exchange anisotropy ($J_x = J_y$) when a strength
of the one quartic Ising-type interaction is fixed ($J_1/J_z = 1.0$) and that of another one 
quartic Ising-type interaction $J_2/J_z$ varies.}
\label{fig2}
\end{figure}

By exploiting the critical condition (\ref{ccef}), let us examine first how a spatial anisotropy 
in two quartic Ising-type interactions influences the critical behavior of the model under investigation. To see this effect, the critical temperature is plotted in figure~\ref{fig2} 
against the anisotropy parameter $J_x/J_z$ for the particular case of the $XXZ$ exchange anisotropy (i.e. $J_x=J_y$) when a strength of the stronger quartic interaction is fixed ($J_1/J_z = 1.0$) 
and a strength of the weaker quartic interaction varies. As one can see, the phase diagram consists 
of two marked wings of critical lines that separate both spontaneously long-range ordered phases 
$|{\rm I} \rangle$ and $|{\rm II} \rangle$, since both critical lines merge together at the ground-state boundary $J_x = J_y = J_z$ between these two phases. It is noteworthy that the left (right) wing represents a line of critical points of the phase $|{\rm I} \rangle$ 
(the phase $|{\rm II} \rangle$), which can be obtained as a numerical solution of the critical condition (\ref{ccef}) by assuming the plus (minus) sign therein. It becomes quite evident 
from figure~\ref{fig2} that one may also found reentrant phase transitions slightly above 
the ground-state boundary between the phases $|{\rm I} \rangle$ and $|{\rm II} \rangle$, 
i.e. under the constraint $J_x = J_y \gtrsim J_z$. In this parameter region, the reentrant phase transitions from the spontaneously ordered phase $|{\rm II} \rangle$ to the disordered phase and 
from the disordered phase to the spontaneously ordered phase $|{\rm I} \rangle$ take place 
because the free energy of the phase $|{\rm I} \rangle$ decreases much more rapidly with the 
increase of temperature than the free energy of the phase $|{\rm II} \rangle$ due to the much 
higher entropy gain of the phase $|{\rm I} \rangle$. Another interesting fact to observe 
here is that a spatial anisotropy in the quartic Ising-type interactions does not qualitatively 
affect the critical behavior of the investigated model system. As it can be easily understood 
from Eqs.~(\ref{ev}) and (\ref{bw5})--(\ref{bw7}), the spatial anisotropy in the quartic 
Ising-type interactions $J_1 \neq J_2$ merely causes an anisotropy in the effective pair 
interactions $R_1 \neq R_2$ of two Ising square lattice coupled together through the effective 
quartic interaction $R_4$ when speaking in the language of the Ising spin representation 
of the corresponding zero-field eight-vertex model. Of course, it is well known that 
the anisotropy in those pair interactions does not fundamentally affect the overall 
critical behavior of the zero-field eight-vertex model.
   
\begin{figure}
\vspace{-0.3cm}
\includegraphics[width=0.5\textwidth]{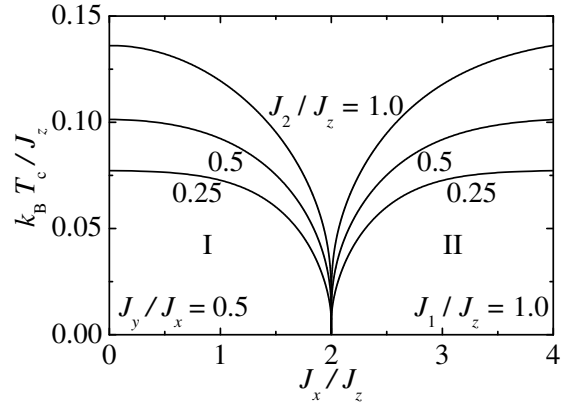}
\vspace{-1.2cm}
\caption{The dimensionless critical temperature as a function of the anisotropy parameter $J_x/J_z$
for the Ising-Heisenberg model with the $XYZ$ exchange anisotropy ($J_x \neq J_y$) when a strength
of the one quartic Ising-type interaction is fixed ($J_1/J_z = 1.0$) and that of another one 
quartic Ising-type interaction $J_2/J_z$ varies.}
\label{fig3}
\end{figure}

\begin{figure}
\includegraphics[width=0.5\textwidth]{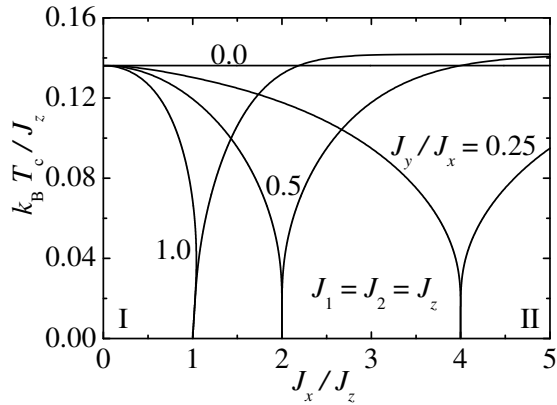}
\vspace{-1.1cm}
\caption{The dimensionless critical temperature as a function of the anisotropy parameter $J_x/J_z$ 
for the spin-1/2 Ising-Heisenberg model with a fixed strength of the quartic Ising-type interactions 
($J_1/J_z = J_2/J_z = 1.0$) at four different values of the exchange anisotropy $J_y/J_x$.}
\label{fig4}
\end{figure}

For the sake of comparison, we have displayed in figure~\ref{fig3} the critical temperature as 
a function of the exchange anisotropy $J_x/J_z$ for one illustrative example of the less symmetric $XYZ$ exchange anisotropy ($J_y / J_x = 0.5$) when a strength of the stronger quartic interaction 
is fixed ($J_1/J_z = 1.0$) and a strength of the weaker quartic interaction varies. It can be 
clearly seen from this figure that the finite-temperature phase diagram of this more general 
case quite closely resembles the one formerly discussed by the analysis of the particular case 
with the $XXZ$ exchange anisotropy. Actually, one still finds two wings of the critical lines 
that separate both spontaneously long-range ordered phases $|{\rm I} \rangle$ and $|{\rm II} \rangle$, whereas the zero-temperature transition between these phases moves towards the higher values of the exchange anisotropy $J_x/J_z$ as it occurs just if the weaker exchange interaction in the $xy$-plane 
$J_y$ overwhelms the one $J_z$ along the quantization axis. However, the most obvious difference between the phase diagrams shown in figures \ref{fig2} and \ref{fig3} is that the latter phase 
diagram does not imply an existence of the reentrant phase transitions near the ground-state 
boundary between the phases $|{\rm I} \rangle$ and $|{\rm II} \rangle$. This specific feature 
can be attributed to the fact that the macroscopical degeneracy of the phase $|{\rm I} \rangle$ 
is entirely lifted whenever the most general case of the $XYZ$ exchange anisotropy is assumed. 
Hence, it follows that both wings of critical lines tend to zero temperature with an infinite 
gradient and thus, there does not emerge reentrant phase transitions near the ground-state
boundary between the phases $|{\rm I} \rangle$ and $|{\rm II} \rangle$. 
To provide an independent check of the aforementioned scenario, 
the critical temperature is plotted in figure \ref{fig4} against the exchange anisotropy $J_x/J_z$ 
for one selected value of the quartic Ising-type interactions ($J_1/J_z = J_2/J_z = 1.0$) 
at four different strengths of the exchange anisotropy $J_y/J_x$. Figure \ref{fig4} apparently 
supports all previous statements, since the critical lines obviously have vertical tangents 
at the zero-temperature transition between the phases $|{\rm I} \rangle$ and $|{\rm II} \rangle$, 
which is generally shifted towards the higher values of the anisotropy parameter $J_x/J_z$ upon strengthening the exchange anisotropy in the $xy$-plane (i.e. when the ratio $J_y/J_x$ is lowered).  
Altogether, it could be concluded that the spatial anisotropy in the quartic Ising-type interactions
has a less significant impact on the overall critical behavior compared to the exchange anisotropy 
of the Heisenberg pair interaction. Therefore, our subsequent analysis will be mainly concentrated 
on how the exchange anisotropy in the Heisenberg pair interaction influences the most essential 
features of the critical behavior. 

\begin{figure}
\includegraphics[width=0.8\textwidth]{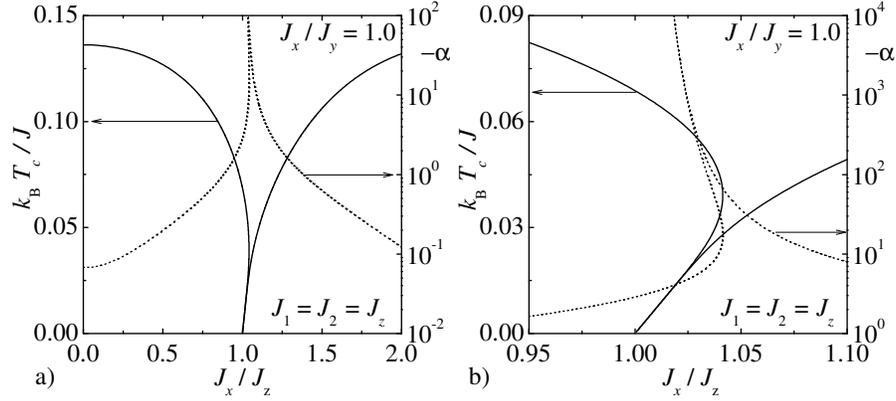}
\vspace{-1.2cm}
\caption{The changes of the critical exponent $\alpha$ along the critical line of the spin-1/2 Ising-Heisenberg model with a fixed relative strength of the quartic Ising-type interactions 
$J_1/J_z = J_2/J_z = 1.0$ and the exchange anisotropy  $J_y/J_x = 1.0$. The solid line, which 
is scaled with respect to the left axis, shows the critical temperature as a function of 
the exchange anisotropy $J_x/J_z$. The broken line, which is scaled with respect to the right 
axis, displays in a semilogarithmic scale the relevant changes of the critical exponent $\alpha$ 
along this critical line. Figure 5b) shows a detail of the phase diagram near 
the zero-temperature transition.}
\label{fig5}
\end{figure}

\begin{figure}
\hspace{-0.7cm}
\includegraphics[width=0.8\textwidth]{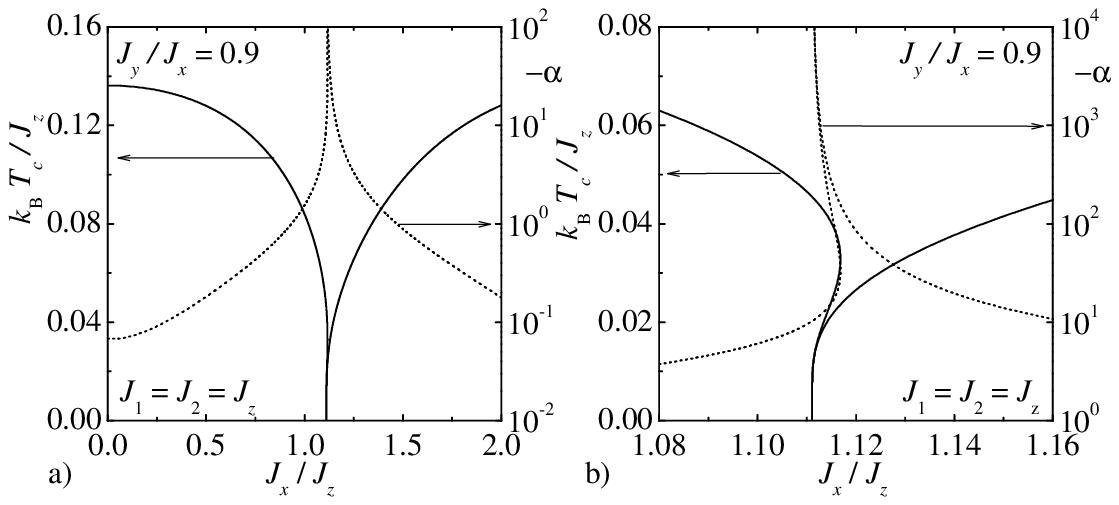}
\vspace{-1.0cm}
\caption{The same as in figure \ref{fig5}, but for the exchange anisotropy $J_y/J_x = 0.9$.}
\label{fig6}
\end{figure}

\begin{figure}
\includegraphics[width=0.8\textwidth]{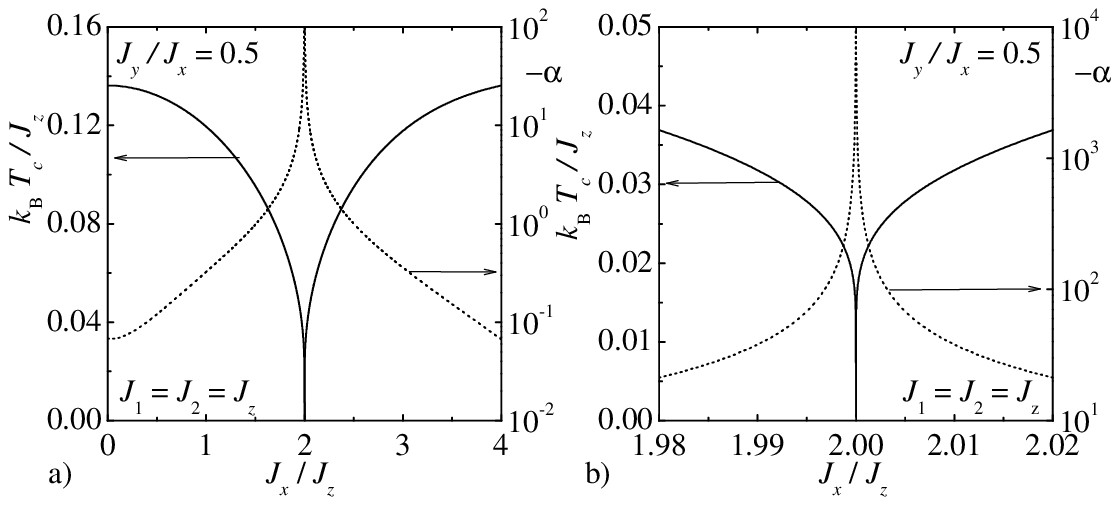}
\vspace{-1.1cm}
\caption{The same as in figure \ref{fig5}, but for the exchange anisotropy $J_y/J_x = 0.5$.}
\label{fig7}
\end{figure}

At this place, let us make a few remarks on possible changes of the critical behavior and 
critical exponents, which can be induced upon varying the anisotropy parameters $J_x/J_z$ and $J_y/J_x$. For this purpose, the figures \ref{fig5}--\ref{fig7} display typical changes of 
the critical temperature and the critical exponent $\alpha$ with the anisotropy parameter 
$J_x/J_z$ for the Ising-Heisenberg model with a fixed strength of the quartic Ising-type 
interactions ($J_1/J_z = J_2/J_z = 1.0$) at three different values of the exchange anisotropy 
$J_y/J_x = 1.0$, $0.9$ and $0.5$, respectively. In these figures, solid lines scaled with respect to left axes depict a variation of the critical temperature with the exchange anisotropy $J_x/J_z$, while 
broken lines scaled with respect to right axes show in a semilogarithmic scale the relevant changes 
of the critical exponent $\alpha$. The figures on the left show the overall finite-temperature 
phase diagrams, whereas the figures on the right show in an enlargened scale the most striking 
part of these phase diagrams in a close vicinity of the ground-state boundary between 
the phases $|{\rm I} \rangle$ and $|{\rm II} \rangle$. 

It is quite obvious from figures \ref{fig5}--\ref{fig7} that the critical exponent $\alpha$ 
varies continuously along the line of critical points, because its value changes with the 
interaction parameters involved in the Hamiltonian (\ref{ham}) according to the relation (\ref{ce})  through the respective changes of the parameter $\mu$. It should be pointed out, moreover, 
that the relevant changes of the critical exponent bring also insight into a nature of phase transitions as the order of phase transitions is proportional to $r=2-\alpha$ (see for instance pp.~16--17 in Reference \cite{baxt82}). From this point of view, the displayed critical lines 
turn out to be lines of rather smooth continuous phase transitions, since the respective variations 
of the critical exponent $\alpha$ are restricted to the range $\alpha \in (-\infty, 0)$ and consequently, the order of phase transitions is $r>2$. However, the most striking finding 
to emerge from figures \ref{fig5}--\ref{fig7} is that the critical exponent $\alpha$ exhibits 
a very special singular behavior in a neighborhood of the ground-state boundary between 
the phases $|{\rm I} \rangle$ and $|{\rm II} \rangle$ where $\alpha \to - \infty$. In this respect, 
the zero-temperature phase transition between the phases $|{\rm I} \rangle$ and $|{\rm II} \rangle$ might be regarded as a phase transition of the infinite order and hence, this special critical point 
in fact represents a quite remarkable quantum critical point. Another interesting finding stems 
from a direct comparison of the figures \ref{fig5}b)--\ref{fig7}b). In agreement with the aforedescribed analysis, the reentrant phase transitions occur in a vicinity of the zero-temperature 
transition between the phases $|{\rm I} \rangle$ and $|{\rm II} \rangle$ by considering the $XXZ$ 
exchange anisotropy (figure \ref{fig5}b), while the reentrant phenomenon obviously vanishes when considering the sufficiently strong $XYZ$ exchange anisotropy (figure \ref{fig7}b). A disappearance 
of the reentrant transitions occurs on behalf of the $XYZ$ exchange anisotropy, which generally 
lifts a macroscopical degeneracy of the phase $|{\rm I} \rangle$, as it has been reasoned by the ground-state analysis. Owing to this fact, the critical lines of both four-fold degenerate phases $|{\rm I} \rangle$ and $|{\rm II} \rangle$ should always tend to zero temperature with the infinite gradient at the ground-state boundary between them whenever the $XYZ$ exchange anisotropy is considered. The quite interesting situation thus appears if there is a small but non-zero exchange anisotropy between both interactions in the $xy$-plane (see figure \ref{fig6}b). In this particular case, both critical lines meet at the ground-state boundary between the phases $|{\rm I} \rangle$ 
and $|{\rm II} \rangle$ with an infinite gradient but they curve in the same direction at small 
enough temperatures, which gives rise to another type of the reentrant phenomenon \cite{suzu87}. Naturally, the greater a difference between the exchange interactions $J_x$ and $J_y$ is, 
the smaller is a temperature range where the reentrant phenomenon may be observed.

\section{Conclusions}

In this work, the critical properties of the spin-1/2 Ising-Heisenberg model with the pair 
$XYZ$ Heisenberg interaction and two quartic Ising-type interactions have been studied in particular
within the exact mapping technique based on the generalized star-square transformation.
This algebraic transformation establishes an exact mapping equivalence between the proposed Ising-Heisenberg model and the Baxter's zero-field (symmetric) eight-vertex model, 
which has been subsequently used for obtaining several interesting exact results for 
the ground-state and finite-temperature phase diagrams, phase transitions and critical phenomena. 
The most interesting finding to emerge from the present study is an exact evidence of 
the quantum critical point, which appears whenever the weaker exchange interaction in 
the $xy$-plane equals the exchange interaction $J_z$ along the quantization axis, 
i.e. whenever $J_z = {\rm inf} \{J_x, J_y \}$. It turns out that the critical exponents exhibit 
a peculiar singular behavior in a close vicinity of this quantum critical point, 
which consequently represents a phase transition of the infinite order.  

The main emphasis of the present paper was to provide a deeper insight into how a spatial 
anisotropy in two quartic Ising-type interactions and the exchange anisotropy in the $XYZ$ 
Heisenberg pair interaction influence the striking weak-universal critical behavior. 
It has been shown that a spatial anisotropy in two quartic Ising-type interactions has less 
pronounced effect on the critical behavior, which changes quantitatively rather than qualitatively 
upon varying the anisotropy in two quartic Ising-type interactions. On the other hand, 
the exchange anisotropy in the $XYZ$ Heisenberg pair interaction significantly changes 
the overall critical behavior, namely, it enables to change a location of the quantum critical 
point as well as an appearance (or disappearance) of the reentrant phase transitions.    

\begin{theacknowledgments}
This work was supported by the Slovak Research and Development Agency under the contract 
LPP-0107-06. The financial support provided by Ministry of Education of Slovak Republic 
under the grant No.~VEGA~1/0128/08 is also gratefully acknowledged.
\end{theacknowledgments}

\end{document}